\documentclass{mem}
\usepackage{natbib}\usepackage{txfonts}\usepackage{balance}
\usepackage{graphicx}
\idline{75}{282}
\begin{document}

\newcommand{\finv}{{F_{\rm inv}}}
\newcommand{\tn}{{t_{\rm nuclear}}}
\newcommand{\thn}{{\thinspace}}
\newcommand{\el}[2]{\ensuremath{^{#1}\mathrm{#2}}}
\newcommand{\Mo}{\rm{M}_\odot}
\newcommand{\Lo}{\rm{L}_\odot}
\newcommand{\Ro}{\rm{R}_\odot}
\def\mnras{MNRAS}
\def\araa{ARA\&A}
\def\apj{Ap.J.}
\def\aap{AAP}
\def\aj{Aj}
\title{
Observational Constraints for $\delta \mu$
Mixing
}

   \subtitle{}

\author{
G.~C. \,Angelou\inst{1},  J.~C. \, Lattanzio\inst{1}, R.~P. \, Church\inst{1,2},
R.~J. \, Stancliffe\inst{1},}

  \offprints{G.C. Angelou}

\institute{
Centre for Stellar and Planetary Astrophysics,  School of Mathematical Sciences,
 Monash University,  Melbourne,  VIC 3800,  Australia
\and
Lund Observatory, Box 43, SE-221 00 Lund, Sweden. 
\newline \email{George.Angelou@monash.edu}
}

\authorrunning{Angelou, et al }

\titlerunning{Constraining $\delta \mu$}

\abstract{We provide a brief review of thermohaline physics and
why it is a candidate extra mixing mechanism during the red giant branch (RGB). 
We discuss how thermohaline mixing (also called $\delta \mu$ mixing) during
the
RGB due to \el{3}{He} burning, is more complicated than the operation of
thermohaline mixing in other stellar contexts (such as following accretion from
a binary companion). We try
to use
observations of carbon depletion in globular clusters to help constrain the
formalism and the diffusion coefficient or mixing velocity that should be used
in stellar models. We are able to match the spread of carbon depletion for metal
poor field giants but are unable to do so for cluster giants, which may show
evidence of mixing prior to even the first dredge-up event.
}
\maketitle{}

\section{Introduction}

The need for extra mixing on the RGB is observationally well
established. Any mechanism (or the combined effect of multiple mechanisms) must
meet the following requirements:
\begin{enumerate}
 \item It must occur after the luminosity bump and continue to operate until
near the tip of the RGB. \citep{1991ApJ...371..578G, 2004MmSAI..75..347W,
2003PASP..115.1211S, 2008AJ....136.2522M}
\item It must occur over a range of masses and metallicities.
\citep{2009A&A...502..267S}
 \item It must deplete \el{7}{Li}. \citep{1998A&A...332..204C,
2009A&A...502..267S}
 \item It must deplete \el{3}{He}. \citep{1986ApJ...302...35D,
1995PhRvL..75.3977H, 1996ApJ...465..887D} 
 \item It must lower the \el{12}{C}/\el{13}{C} ratio \citep{1994A&A...282..811C,
1996ASPC...98..213C} 
 \item It must deplete the carbon abundance and increase the nitrogen abundance.
 \citep{2009A&A...502..267S, 2003PASP..115.1211S, 2008AJ....136.2522M}
\end{enumerate}

These criteria suggest that, in order for theory to remain consistent with
observations, material must be mixed through radiative regions, processed by the
H-shell, and mixed back into the enevelope.
This requirement is often referred to as deep mixing because mixing
deeper than the formal convective boundary into the radiative zones will lead to
material being exposed
to regions of higher temperature and will result in the required additional
processing. In general
the \el{12}{C}/\el{13}{C} ratio is
used to probe the efficiency of first dredge up (FDU);
\citep{1975MNRAS.170P...7D,1976ApJ...210..694T} and is also used as a tracer of
the extent of deep mixing. A good example of this was
\citet{1979ApJ...229..624S} who were the first to use the isotopes to
investigate the role of rotational mixing on the RGB. More recently
\citet{2006A&A...453..261P} have shown that, whilst rotation does reduce
the \el{12}{C}/\el{13}{C} isotopic ratio, it is unable to explain the values
seen
in giant photospheres. Although it is understood that extra  mixing must take
place, only recently has a
mechanism (i.e. thermohaline mixing) been discovered that can potentially
satisfy all of the aforementioned criteria \citep{2006Sci...314.1580E}.

\section{Thermohaline Mixing In Stars}
\subsection{Historical Overview}
Thermohaline mixing was first studied in the Earth's oceans by \citet{stern}
where stratified
warm salty water sits upon a cool unsalted layer. The layers are initially stable. However, heat diffuses more
quickly than composition so the warmer layers cool. Now they are
simply denser than the material underneath, and a turnover is
initiated via the formation of lengthy ``fingers" of cooler salty
water reaching down into the cold fresh water. This displaces cool
fresh water upwards, and a mixing occurs. On a slower timescale the
salt diffuses out of the salty cool water to reach a new saltiness in
the mixed region.

\indent This double diffusive mixing was first applied to a stellar
context by \citet{1969ApJ...157..673S}. \citet{1972ApJ...172..165U} applied this
to a perfect gas and \citet{1980A&A....91..175K} extended this to allow for a
non-perfect gas which included radiation pressure and degeneracy. There were two
obvious situations in which they applied thermohaline mixing. Firstly, during
pre-main sequence contraction when in-situ \el{3}{He} burning lowers the local
mean
molecular weight, $\mu$ because the reaction
\begin{equation} \label{eqn:He3}
  \el{3}{He}\left( \el{3}{He},2 \rm{p}\right) \el{4}{He}
   \end{equation}
produces more particles than it destroys.
The mixing is determined by the competition of the heat diffusion and the
difference in composition but it is driven by the change in local molecular
weight. This was found to have little effect, due to the short
pre-main-sequence time scale and the fact the star becomes fully convective
before reaching the Zero Age Main Sequence (ZAMS). The second case consisdered
was during
the core flash, when during off-centre He ignition, carbon-rich material sits
upon helium-rich material. This also was considered to have little effect on the
evolution primarly due to the uncertainty of competing timescales. 
The mixing must occur before the star settles down to quiescent helium burning.
\citet{2006Sci...314.1580E} also showed 
that a small amount of overhooting inwards could remove the molecular weight
inversion on a dynamical timescale.  
 
\subsection{Application to the RGB}
\citet{2006Sci...314.1580E} used a 3D hydrodynamical stellar code \citep{2006ApJ...639..405D} to show an instability develops due to \el{3}{He} burning along the RGB, an instability that \citet{2007A&A...467L..15C} 
model as thermohaline mixing. \citet{2007A&A...467L..15C} and \citet{2008ApJ...677..581E} have modelled this \el{3}{He} burning driven instability in their 1D codes and demonstrated the significant  effect the mixing has during the RGB. 
Following FDU, the convective envelope recedes, leaving
behind a homogeneous region. Any composition and molecular weight gradient has
been removed due to the convective mixing. As the hydrogen burning shell begins
to advance, \el{3}{He} begins to burn.
From Equation \ref{eqn:He3}  it can be seen that this reaction creates a local
molecular weight inversion; \citet{2008ApJ...677..581E} found its magnitude to
be 
of the order ${\bigtriangleup \mu}/{\mu}$ $\sim$ 10$^{-4}$. Although the
inversion seems small, convection is in fact driven by a similarly small
superadiabaticity. Usually such a small change in the local molecular
weight would have almost no effect, as it would be swamped by the existing
$\mu$ gradient produced by the burning of other species. It is this unique
situation where \el{3}{He} begins to burn before the other species and the fact
that
first dredge up has homogenised the region that allows the
inversion to develop.
\newline \indent Although there is no salt in the star the process is doubly diffusive and thus labelled
thermohaline mixing. The authors refer to this as $\delta \mu$ mixing
to emphasise that the mechanism that drives the mixing and the fact it is more
complex than the other examples of thermohaline mixing. As \el{3}{He} burns, a
parcel forms that is hotter and has lower molecular weight than its
surroundings. It quickly expands (and begins to cool) in order to establish
pressure equilibrium. The expansion reduces the density and therefore the
element becomes buoyant. The parcel rises until it finds an equilibrium
point where the external pressure and density are equal to that inside the
bubble.
This is expected to be a small displacement which occurs on a dynamical
timescale. 

\indent As the molecular weight inside the bubble is lower than its surroundings
the
equilibrium point must correspond to a place where the external temperature is
higher than that of the bubble. The temperature inside the bubble will be lower
than its surroundings: 

\begin{equation} 
 \rm{ \frac{T_i}{T_o}=\frac{ \mu_i}{ \mu_o}},
\end{equation}
where subscript i denotes the inside of the bubble and subscript o denotes the
surroundings.
 As heat begins to diffuse into the parcel, we expect layers will start to strip
off
in the
form of long fingers. It is this
secondary mixing that governs the overall mixing timescale. The mixing
cycles in fresh \el{3}{He} from the envelope reservoir, while CN-processed
material is cycled into the convection zone. 
\newline \indent \citet[][EDL hereafter]{2008ApJ...677..581E}  found that this
mixing satisfies
the criteria outlined in Section 1. The level of depletion of the carbon
isotopes will
depend on the efficiency of the mechanism. EDL estimated
the mixing speed and with their formula for the diffusion coefficient found that
a window of
three orders of magnitude in the mixing velocity can lead to observed
levels of \el{12}{C}/\el{13}{C} and \el{3}{He} depletion.
\citet{1972ApJ...172..165U} and \cite*{1980A&A....91..175K}  use essentially the
same formula for the diffusion coefficient (UKRT hereafter) but their geometric coefficients
vary by two
orders of magnitude. \citet{2007A&A...467L..15C} have applied the UKRT mixing to
the RGB. Both \citet{2007A&A...467L..15C} and EDL see the \el{12}{C}/\el{13}{C} ratio is reduced to similar levels.
In this study we will attempt to use globular cluster observations to constrain
both the form of the diffusion coefficient and the mixing velocity. 

\section{The Mixing Speed}
In order to implement $\delta \mu$ mixing into our 1D codes we must consider the
following:
\begin{enumerate}
 \item Which formalism should be used? Here we will limit our investigation to
the EDL and UKRT prescriptions for the diffusion coefficient. 
 \item Once the preferred formalism is identified what mixing velocity is needed
to match observations? What values do we use for any free parameters?
 \item The \el{12}{C}/\el{13}{C} ratio is generally used as a proxy to probe the
extent of mixing. This quickly saturates in low metallicity stars and therefore
could be misleading. Is there a better way to try to constrain the velocity?  
\end{enumerate}

\indent EDL postulated the following formula based
on the velocities from their 3D code in analogy with the existing convective
formalism in their code:
\begin{equation}
 D=\left\{\begin{array}{ccc}\frac{\finv r^2}{\tn}\thn(\mu - \mu_{\rm
min})&\mbox{ if }
(k\ge k_{\rm min})\\\\0 &\mbox{ if }(k\le k_{\rm min}) , \end{array} \right. 
\end{equation}

where  $\mu_{\rm{min}}$ is the smallest value of $\mu$ in the current
model, $k$ the mesh point number, counted outwards from the centre, $r$ is the
radial coordinate, $F_{\rm{inv}}$ is a constant  which is selected to obtain the
desired
mixing efficiency and $\tn$ is an estimate of the nuclear evolution timescale
(see EDL).
 
 This formulation ensured the correct region was mixed but ensures the mixing is formally zero at the  position where $\mu$ has
its minimum even though
it should presumably  be the most efficient at this point. EDL give upper
and lower estimates for the mixing velocity and find that they can alter the
speed by three
orders of magnitude and still produce the
observed levels of \el{12}{C}/\el{13}{C} and \el{3}{He}. 
 \citet{2007A&A...467L..15C} adopt the UKRT formula 
\begin{equation} \label{eqn:kip}
D_t =  C_t \,  K  \left({\varphi \over \delta}\right){- \nabla_\mu \over
(\nabla_{\rm ad} - \nabla)} \quad \hbox{for} \;  \nabla_\mu < 0 ,
\label{dt}
\end{equation}
where  $\nabla = (\partial \ln T / \partial \ln P)$, $\varphi = (\partial \ln
\rho / \partial \ln \mu)_{P,T}$, $\delta=-(\partial \ln \rho / \partial \ln
T)_{P,\mu}$ and $C_t$ is the geometric factor.

Empirical studies of fluids in laboratory conditions led \citet{1972ApJ...172..165U}  to
determine that  $C_{\rm{t}}$  $\sim$ 1000. He saw the development of long salt
fingers with lengths that were larger than their
diameters, which led to efficient mixing. Kippenhahn on the other hand
envisaged the classical picture where mixing is due to blobs and thus determined
 $C_{\rm{t}}$ $\sim$ 10.  

 \indent We have run stellar models of various masses, with  both EDL and
UKRT mixing. We tested different values of $F_{\rm{inv}}$ and  $C_{\rm{t}}$
 in order to alter the efficiency of mixing.
To test our models for the extra mixing we chose to use the carbon abundance as
a function of $M_{\rm{V}}$ as determined by \citet{2003PASP..115.1211S}. They
plotted carbon abundance as a
function of visual magnitude for a variety of globular clusters. In doing so
they were able to clearly demonstrate the depletion of carbon along the RGB.
Globular cluster have always been an excellent test bed for stellar theory and
by trying to match the carbon depletion for various red giant branches we have
an alternative abundance test for mixing efficiency.     
\section{Results}

 \begin{figure*}[ht]
\resizebox{\hsize}{!}{\includegraphics[scale=.6]{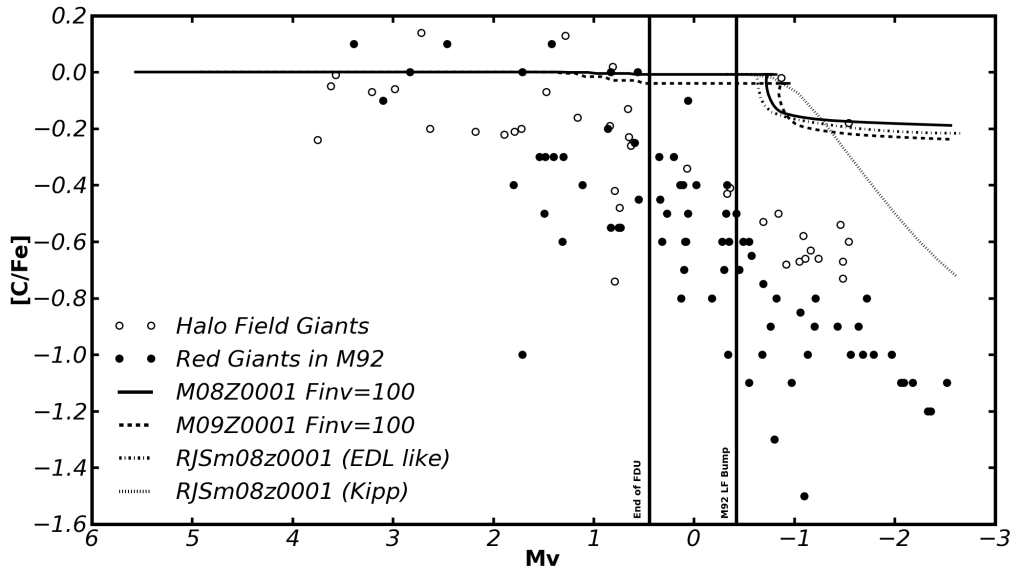}}
\caption{\footnotesize
We plot the carbon abundance [C/Fe] as a function of $M_{\rm{V}}$ for observed giants
and our models. We include cluster giants from M92 (solid circles) along with
Galactic
halo field giants (open circles) where the metallicity covers [Fe/H] =-1 to -2. Both sets of observational data were taken from\citet{2003PASP..115.1211S}. 
 The solid line and the bashed line correspond to models run with
MONSTAR. The solid black line is the
evolutionary track for a 0.8 $\Mo$ star up until the core flash. The dashed line
is the evolution of a 0.9 $\Mo$ star. The dotted and dot-dashed lines are
models run with the
Eggleton code which has been modified by \citet{2009MNRAS.396.1699S}. The
dot-dashed
line corresponds to the evolution of a 0.8 $\Mo$ star using an EDL formalism.
The dotted line is for a 0.8 $\Mo$ star using a UKRT style mixing and
 $C_{\rm{t}}$ = 1000.} 
\end{figure*}

In Figure 1 we plot carbon abundance for stars in the Galactic globular cluster
M92 and the Galactic
halo from \citet{2003PASP..115.1211S}. Open circles denote galactic field giants
whose metallicity ranges from -2.0 $\leq$ [Fe/H] $\leq$ -1.0 also taken from \citet{2003PASP..115.1211S} . The filled circles
correspond to RGB stars in M92. In both the field and the halo it is immediately
obvious that there is carbon depletion as stars ascend the giant branch. If our
models are able to match the carbon depletion we may be able to constrain the
thermohaline
mixing formalism and velocity. Another thing to notice before turning to the
models is the spread in carbon for a given visual magnitude. 
We attribute this to the primordial abundances of the cluster with the most C-rich at a given magnitude being the ``normal'' stars. 
The spread in C at a given magnitude is assumed to be of primordial origin as is the case with many other globular clusters.
Our primary aim is to match the level of
carbon depletion. That is, we are concerned with matching the decrease in the
upper and lower limits of the [C/Fe] values, as a function of
magnitude. The solid and dashed lines were computed using MONSTAR 
(\citealt{2008A&A...490..769C}). We have evolved
a 0.8 $\Mo$ and a 0.9 $\Mo$ star until the core flash. These masses straddle the
age limits of stars in this cluster. A metallicity of Z=0.0001 was used to match
that of the M92 where [Fe/H]=-2.2 \citep{2001PASP..113..326B}. The EDL mixing
quickly destroys the \el{3}{He} without significantly altering the FDU values of
carbon. We believe this model is not mixing to high enough temperatures. As
mentioned in the previous section, the mixing speed is formally zero at the
position where $\mu$ has its minimum. 
By not mixing at the minimum properly the $\mu$ profile is being affected and
carbon is
not being
exposed to the required temperature in either model. 
\newline \indent
The dot-dashed line is a model computed using the Eggleton code,
\citep{1971MNRAS.151..351E, 2009MNRAS.396.1699S}. It too is of mass 0.8 $\Mo$
and
corresponds to the metallicity of M92 however it is run without mass loss.
Running without mass loss here will result in less carbon depletion than we
would otherwise expect, therefore will serve as a lower limit for the depletion of carbon.  An EDL style formula for the diffusion coefficient 
is used in this calculation, that is there is a dependence on the
position where $\mu$ reaches it's minimum.
The $F_{\rm{inv}}$ here was cailbrated such that a 1.5 $\Mo$, Z=0.0001 model
gave the same level
of carbon depletion on the RGB as a 1.5 $\Mo$ Z=0.0001 model with UKRT
mixing where
 $C_{\rm{t}}$=1000, (see \citealt{2010MNRAS.403..505S} for more detail). 

 \indent
The dotted line is a model with a UKRT prescription taken
from \citet{2009MNRAS.396.2313S} where  $C_{\rm{t}}$=1000. This also was run without
mass loss.  The UKRT mixing is a
local formalism
that is dependent on the $\mu$ gradient. Unlike in the
EDL case this translates to the mixing being more efficient at the position
where  $\mu$ reaches its minimum.
In both cases carbon is brought down from the envelope but here 
it is mixed to the position of lowest molecular weight  and hence exposed to the
shell much faster.
The high temperature gradient ensures that mixing only a little deeper will
see the carbon undergo larger depletion. This is of course all dependent on the
amount of \el{3}{He} available to drive the mixing. We see that the UKRT
mixing can lead to levels of depletion seen in the field giants. Given that the
field giants and M92 stars are of similar age and metallicity it is interesting
the cluster stars undergo more substantial depletion. We defer the discussion of
why this is to subsequent work.

\section{Conclusion}
Our initial motivation behind this paper was to use
the observed variation of carbon abundances on the giant
branch to help constrain some of the uncertainties
present in the thermohaline mixing which
we believe is operating during the red-giant phase.
Drawn by the best data being available for M92, we
chose this as our first attempt to fit the observations.
The fact that we have failed in our aim has nevertheless
taught us three important things:
\begin{enumerate}
 \item  The functional form of the diffusion co-efficient
strongly influences the depletion of carbon.
\item Comparing the carbon isotope ratio
is not necessarily useful because it saturates at the equilibrium
value of about four while C continues to burn into N.
\item The carbon abundances in M92 may provide a very serious
challenge for  stellar evolution, independent of
any deep-mixing mechanism. At the same time it could in fact be telling us
something very important about the deep mixing process. 
\end{enumerate}
Concerning the first point we note that the simple formula
used by EDL causes an initially rapid depletion and then
a leveling off, which does not seem to match the observations
for metal-poor globular clusters. The UKRT description
results in a more gradual depletion and may be a better description.
Neither depletes the carbon by enough to match the observations,
but we note that the Eggleton models here are run without mass loss which will
exaggerate the discrepancy. 

We believe that the third point is more fundamental. The data
for M92 clearly show depletion in [C/Fe] for stars with
magnitudes $M_{\rm{V}}$ $>$ 1. Note that standard stellar evolution
predicts that the first dredge-up does not produce observable
abundance changes for these stars, and that this dredge-up
does not finish until a magnitude $M_{\rm{V}}$ $\sim$ +0.5. By this stage
in the evolution, the stars are already showing depletions
of C of order 0.5 dex. Further, the bump in the luminosity
function (hereafter LF bump) is observed to be at $M_V$ $\sim -0.4$
\citep{1990A&A...238...95F}. According to the
usual ideas, deep-mixing (by whatever the mechanism) is
inhibited until the star reaches the LF bump
and the advancing H-shell removes the molecular weight
discontinuity left behind by the receding convective envelope
at the end of first dredge-up. In the case of M92 the
stars on the giant branch have already depleted their [C/Fe] by about 0.8
dex when they reach this stage. If we have to postulate that some form of
mixing begins sufficiently early to produce this depletion, then the
mixing
must necessarily remove the abundance discontinuity
that is itself responsible for the observed LF bump!
The resulting contradiction
produces, in our view, a serious problem for stellar astrophysics.

It is worth noting that the LF bump in M92 is not as clearly
visible as it is in more metal-rich clusters. \citet{1990A&A...238...95F}
had to co-add data for three very similar clusters to make it
visible in the data. Indeed, recent work by \citet{2007AJ....133.2787P} provides little
evidence for a bump in the observed LF of M92. These authors
show that even the theoretically predicted bump is small (see
also \citealt{1978IAUS...80..333S}). We are left trying to identify cause and effect:
is the reduced bump the result of a reduced discontinuity in the
molecular weight in this case, which is not enough to prevent mixing
before the disconintuity is erased by nuclear burning? Or does some
mixing begin before the bump is reached, with the necessity of that
mixing reducing the molecular weight discontinuity?
\newline \indent We note that we are not the first to have noticed this problem,
as it has been discussed by (at least) \citet{2008AJ....136.2522M},
\citet{2001PASP..113..326B}, and
\citet{1986PASP...98..473L}. However, the data in Figure 1 are compiled from various sources and
this presents a uniformity problem. Offsets by 0.3 dex are possible (G
Smith, private communication) and could be the cause of the apparent
contradiction. Certainly to use M92 as a constraint for $\delta \mu$
mixing requires a homogeneous set of data covering a wide range of
luminosities. Such data are simply not available at present, but would
prove extremely valuable.

\bibliographystyle{aa.bst}
\bibliography{Torino2}

\end{document}